\documentclass[aps,superscriptaddress,reprint,nofootinbib]{revtex4-2}
\usepackage{lineno}
\usepackage{booktabs}
\usepackage{graphicx}
\usepackage{amsmath}
\usepackage{amssymb}
\usepackage{amsfonts}
\usepackage{filecontents}
\usepackage{color}
\usepackage[normalem]{ulem} 
\usepackage{soul}
\usepackage{epstopdf}
\usepackage{nameref}
\usepackage{hyperref}
\hypersetup{hidelinks}
\usepackage{fancyhdr}
\usepackage{float}
\usepackage{gensymb}
\usepackage{physics}
\usepackage{upgreek}
\newcounter{fig}
\usepackage{lipsum}
\usepackage[usenames,dvipsnames,svgnames,table]{xcolor}
\usepackage{enumitem}
\usepackage{multirow}
\usepackage{tabularx}
\usepackage{array}

\DeclareRobustCommand{\SkipTocEntry}[5]{}

\begin{document}

\title{PT-PINNs: A Parametric Engineering Turbulence Solver based on Physics-Informed Neural Networks}

\author{Liang Jiang}
\affiliation{State Key Laboratory of Clean Energy Utilization, Zhejiang University, Hangzhou, China}

\author{Yuzhou Cheng}
\affiliation{Shanghai Institute for Advanced Study of Zhejiang University, Shanghai, China}

\author{Kun Luo}
\email[Email to: ]{zjulk@zju.edu.cn}
\affiliation{State Key Laboratory of Clean Energy Utilization, Zhejiang University, Hangzhou, China}

\author{Jianren Fan}
\affiliation{State Key Laboratory of Clean Energy Utilization, Zhejiang University, Hangzhou, China}

\begin{abstract}
\textbf{
Physics-informed neural networks (PINNs) demonstrate promising potential in parameterized engineering turbulence optimization problems but face challenges, such as high data requirements and low computational accuracy when applied to engineering turbulence problems. 
This study proposes a framework that enhances the ability of PINNs to solve parametric turbulence problems without training datasets from experiments or CFD-Parametric Turbulence PINNs (PT-PINNs)). 
Two key methods are introduced to improve the accuracy and robustness of this framework. 
The first is a soft constraint method for turbulent viscosity calculation. The second is a pre-training method based on the conservation of flow rate in the flow field.
The effectiveness of PT-PINNs is validated using a three-dimensional backward-facing step (BFS) turbulence problem with two varying parameters ($Re_h$ = 3000-200000, ER = 1.1-1.5).
PT-PINNs produce predictions that closely match experimental data and computational fluid dynamics (CFD) results across various conditions.
Moreover, PT-PINNs offer a computational efficiency advantage over traditional CFD methods.
The total time required to construct the parametric BFS turbulence model is 39 hours, one-sixteenth of the time required by traditional numerical methods.
The inference time for a single-condition prediction is just 40 seconds—only 0.5\% of a single CFD computation.
These findings highlight the potential of PT-PINNs for future applications in engineering turbulence optimization problems.
} 
\end{abstract}

\maketitle

Solving turbulent flow fields plays a critical role in engineering applications such as combustion chamber design \cite{Schmitt_Poinsot_Schuermans_Geigle_2007}, aerodynamic optimization of vehicles \cite{Aerodynamic}, and thermal management of motors and chips \cite{Khalaj-thermal-management}.
Over the past decades, various numerical methods, including the finite difference method (FDM) \cite{MOHANRAI199115}, finite volume method (FVM) \cite{RHIE}, finite element method (FEM) \cite{Bassi}, spectral method \cite{Orszag_Patera_1983}, and Lattice Boltzmann method (LBM) \cite{chikatamarla2010lattice}, have been developed and applied to solve turbulent flow fields. These methods are typically based on discretization techniques and predict flow behavior by numerically solving the Navier-Stokes equations.
However, it is often necessary to parameterize the turbulent flow field in practical engineering design. For problems with varying geometries and boundary conditions, these methods often require multiple simulations.
In such cases, even with the use of methods such as large eddy simulation (LES) \cite{PitschLES} or Reynolds-averaged navier-stokes simulation (RANS) \cite{Spalart}, these numerical methods still consume significant time and energy to solve the flow fields under multiple operating conditions \cite{Aerodynamic}.
At the same time, the vast amounts of data generated from multiple simulations often hinder storage and retrieval, increasing the difficulty of building intelligent digital twin systems for modern design, equipment operation, and maintenance.

In recent years, driven by advancements in artificial intelligence, data-driven methods for parameterized flow field solutions have been developed \cite{GUOCNN, LUIRO, NAKRO}.
These methods rely on existing experimental or computational fluid dynamics (CFD) data, leveraging machine learning techniques to train neural networks that take spatial coordinates, geometry, and operating parameters as inputs and output flow field variables.
The resulting machine learning models are compact, easily storable, and enable fast inference of flow fields under varying parameters.
Operator learning methods such as DeepONet \cite{lu2019deeponet} and Fourier Neural Operators (FNO) \cite{li2020fourier} have recently improved the generalization capabilities of these models through network architecture optimization.
However, these models still require extensive datasets for training, and their predictive capabilities degrade significantly when the parameter range is insufficiently represented in the dataset or when the training data lacks accuracy.
In many engineering applications, acquiring sufficient high-quality data remains costly and challenging.

Data-driven machine learning surrogate models for flow fields are inherently limited in generalization accuracy and physical interpretability. As a result, physics-driven machine learning methods have recently emerged as a promising paradigm.
Raissi et al. \cite{raissi2019physics} introduced the Physics-Informed Neural Networks (PINNs) method for solving partial differential equations (PDEs). 
Since neural networks are inherently differentiable, automatic differentiation \cite{baydin2018automatic} can be used to compute PDE residuals across the entire input space. 
By incorporating the governing PDEs as loss functions in the neural network training process, functional solutions to PDEs under specific initial and boundary conditions can be obtained.
They successfully applied PINNs to solve the cylinder flow problem, achieving highly accurate predictions of flow fields with minimal data requirements \cite{raissi2019physics}.
Jin et al. \cite{sun2023physics} applied PINNs to solve two-dimensional (2D) laminar flow around airfoil geometries, demonstrating good agreement with CFD results and showcasing the potential of PINNs for parameterized flow field solutions.
Jin et al. \cite{jin2021nsfnets} proposed a PINNs-based incompressible flow solver, NSFnet, which achieved promising results for turbulent channel flows at $Re_\tau = 1000$.

Despite the potential of PINNs in computational fluid dynamics, their application to turbulent flow problems remains challenging due to the inherent complexity of turbulence.
For instance, Pioch et al. \cite{fluids8020043} demonstrated that PINNs failed to accurately predict 2D backward-facing step (BFS) turbulent flows (Re=5100) using 
$k$-$\epsilon$ and $k$-$\omega$ models unless supplemented with direct numerical simulation (DNS) data. 
Even with partial internal data assimilation, significant deviations persisted in flow field predictions. 
Yadav et al. \cite{yadav2024rf} applied PINNs to solve 2D turbulent flame, and found that the model could not predict the velocity fields correctly when only the sparse training data of the temperature field was applied. 
Many other existing studies also reveal critical limitations when applying PINNs to solve turbulence problems without auxiliary data \cite{cai2021physics, ghosh2024usingparametricpinnspredicting}.

Recent attempts to address these limitations have yielded mixed outcomes.
Cao et al. \cite{cao2023efficient} attempted to use PINNs for the parametrization design of 2D duct flow deflectors (Re=$1\times10^5$) and achieved good agreement with CFD results in zero-equation turbulence modeling. However, the method was limited to simplified 2D problems.
Meanwhile, Gafoor et al. \cite{gafoor2025physics} observed persistent discrepancies between PINNs predictions and experimental data in wind turbine wake studies using $k$-$\epsilon$ models.
To enhance PINNs' capability for multiscale turbulence, advanced architectures such as Time-stepping-oriented Neural Networks (TSONN) \cite{cao2025solving, cao2023tsonntimesteppingorientedneuralnetwork}, domain decomposition \cite{Moseley_2023}, and operator network-based approaches \cite{li2023physicsinformedneuraloperatorlearning} have emerged. However, these methods incur substantial computational costs and lack validation for three-dimensional turbulent flows.
Overall, PINNs for solving parameterized engineering flows remain in a preliminary stage, typically limited to laminar flows \cite{LIU2024104937, LIU2023113094,cao2025laminarflowsairfoils} or 2D turbulent flow \cite{cao2023efficient}. Accurately solving three-dimensional high Reynolds number engineering turbulence using two-equation turbulence models with wall functions remains challenging without the assistance of internal flow field data for training.

\begin{figure*}
\refstepcounter{fig}
\includegraphics[width=0.95\linewidth]{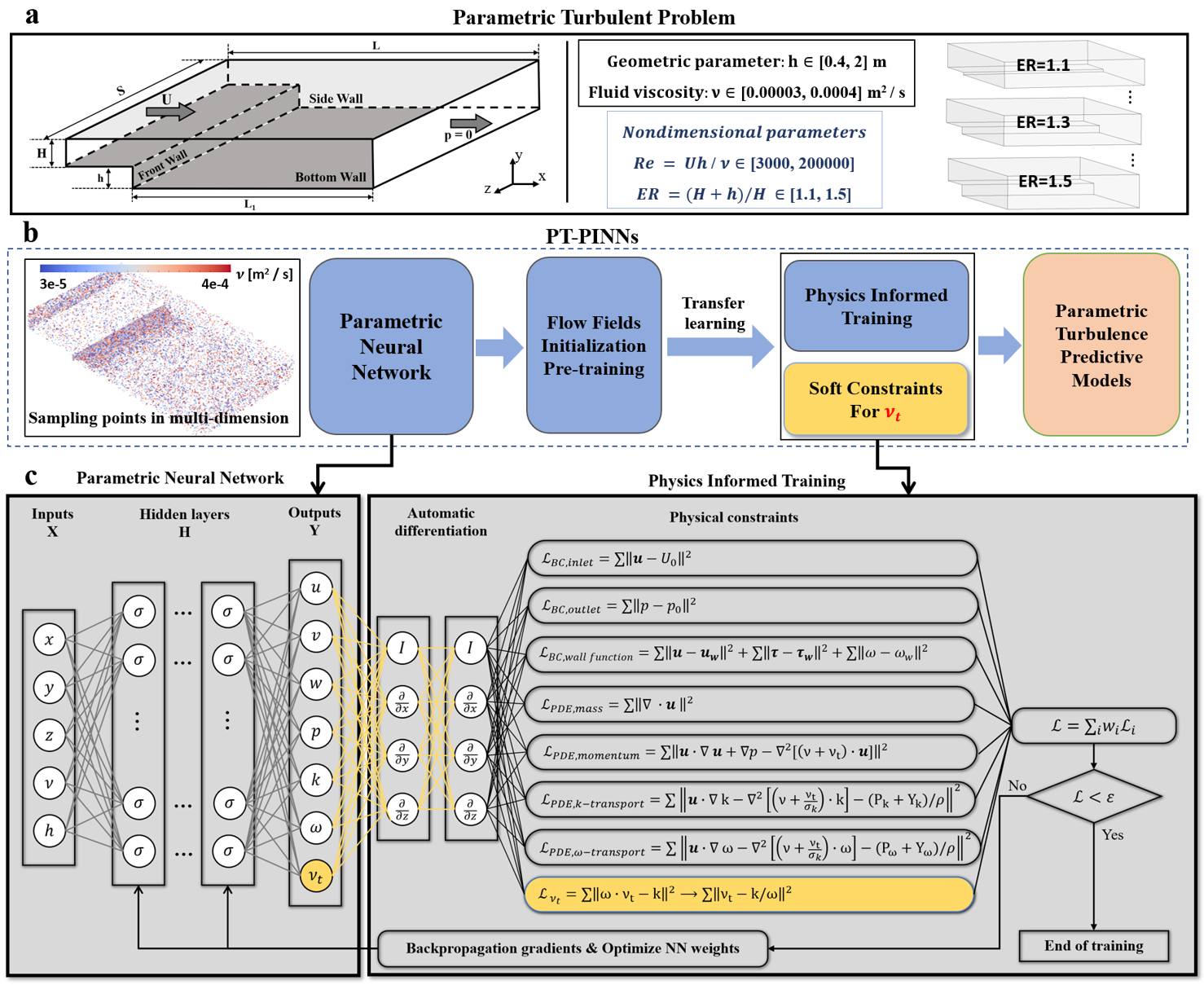}
\caption{\textbf{The framework of the PT-PINNs.} \textbf{a.} The configuration of the parametric 3D turbulent BFS problem. \textbf{b.} The flow chart of PT-PINNs. \textbf{c.} The network structure and the training process of PT-PINNs.  
}
\label{fig_1}
\end{figure*}

\begin{figure*}
\refstepcounter{fig}
\includegraphics[width=0.95\linewidth]{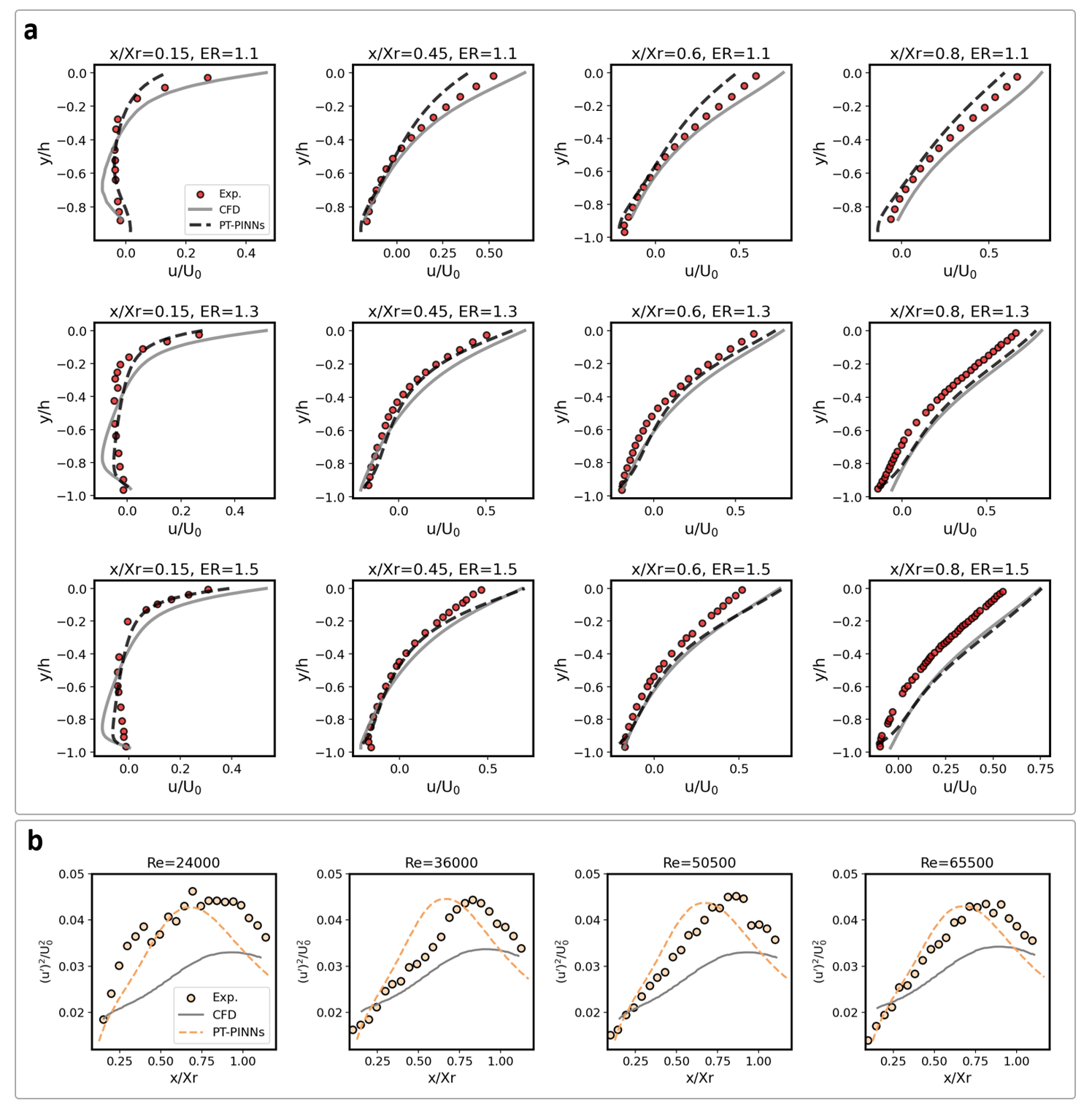}
\caption{\textbf{Quantitative comparisons of PT-PINNs' flow field velocity $\mathrm{u}$ and turbulence stress $\mathrm{(u')^2/U_0^2}$ prediction with experimental \cite{nadge2014high} and CFD results in the recirculation region.} \textbf{a.} Velocity u distribution of the section line at x/Xr=0.15, 0.45. 0.6. 0.85 positions in Z=0 Plane. \textbf{b.} Stream-wise variation of maximum turbulent stress $\mathrm{(u')^2/U_0^2}$ under different Reynolds numbers when ER=1.3. The X axis is normalized by mean reattachment length Xr. 
}
\label{fig_4}
\end{figure*}

In this work, we proposed a parameterized turbulent flow field prediction framework based on PINNs. This framework effectively enhances the predictive capability of PINNs for parameterized steady-state turbulent flow field prediction.
We applied two training methods within this framework. The first is a soft constraint calculation method for turbulent viscosity, which takes turbulent viscosity as the output of the neural network and transforms the algebraic calculation formula of turbulent viscosity in the two-equation turbulence model into the loss function for turbulent viscosity. The second is a pre-training method based on the conservation of flow rate in the flow field.
We validated the framework's effectiveness through a parameterized turbulent backward-facing step flow problem with $Re_{h} \in \left[3000,200,000\right]$, incorporating two variable parameters: the step expansion ratio (ER) and fluid viscosity $\nu$.
The parametric PINNs model was trained solely using the PDEs of k-$\omega$ turbulence model and boundary conditions.
By comparing the predictions of PINNs with CFD results and experimental results from Nadge et al. \cite{nadge2014high} across different geometric shapes and Reynolds numbers, we demonstrated the framework's reliability.
Furthermore, we explored the effects of the two key training methods in this framework. By contrasting the parameterized k-$\omega$ turbulence model for backward-facing step flows under different weight balancing of PDEs and boundary conditions, both with and without these methods, we confirmed their role in enhancing PINNs' ability to balance the losses of multiple complex partial differential equations in turbulence modeling and to avoid convergence to local optima.

\section*{Results}
\addtocontents{toc}{\SkipTocEntry}
\subsection*{Framework Overview}

\begin{figure*}[t]
\refstepcounter{fig}
\includegraphics[width=0.99\linewidth]{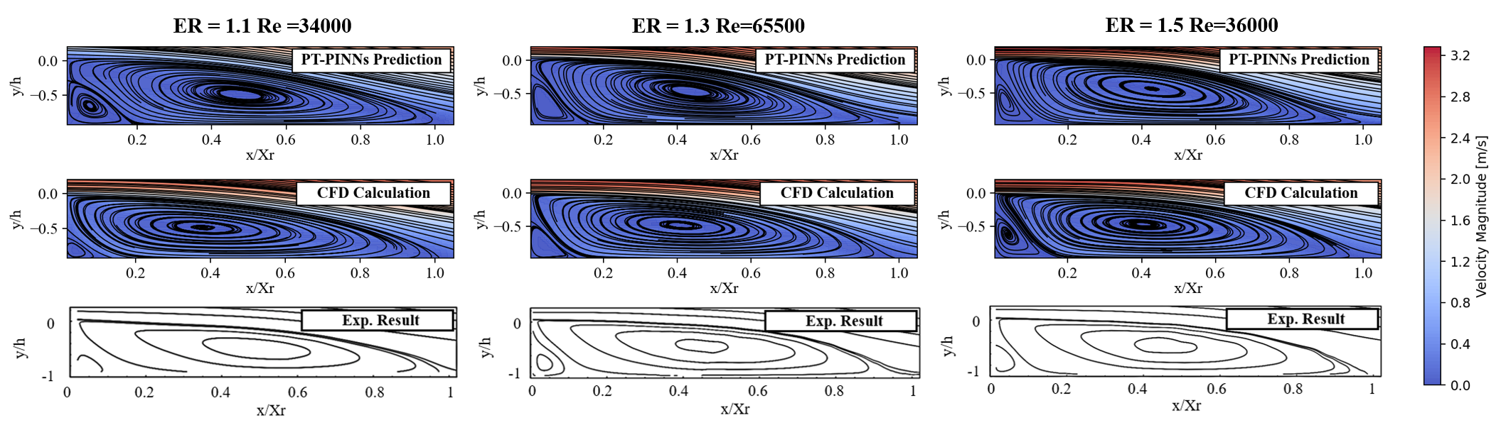}
\caption{\textbf{Comparisons of PT-PINNs predictions of streamlines of the recirculation vortex at the Z=0 Plane with experimental \cite{nadge2014high} and CFD results under different Reynolds numbers and expansion ratios.}  
}
\label{fig_2}
\end{figure*}

A parametric, steady-state turbulent flow field prediction framework based on PINNs is developed in this study. The process for solving a specific turbulent flow field prediction problem using this framework is illustrated in Fig.~\ref{fig_1}.
PT-PINNs directly employ neural networks to solve parameterized, high-dimensional turbulence problems by feeding both spatial parameters and parameters related to geometric shapes, operating conditions, and fluid properties as inputs into the network.
It requires only a single model training session to solve parameterized turbulence problems with multiple variable parameters.

We introduced a pre-training method based on flow field initialization to this framework. The loss function during the pre-training phase consists of parameterized boundary conditions, integral interface constraints, and specified interior numerical constraints based on flow field flux conservation. 
The number of iterations in the pre-training phase is 1/20th of the total training steps. Upon completion of the pre-training, we employed a transfer learning approach to use the pre-trained network as the initial network for the formal training, thereby accelerating the overall convergence and enhancing training stability. The flowchart of the overall framework is shown in Fig.~\ref{fig_1}b.
In addition to flow field variables such as velocity, pressure, and turbulent kinetic energy, we explicitly defined turbulent viscosity as one of the neural network outputs, although turbulent viscosity can be algebraically computed by two-equation or zero-equation turbulence models. 
We transformed the algebraic expressions into loss functions to constrain turbulent viscosity as shown in Fig.~\ref{fig_1}c. This constraint ensures that its effect is nearly equivalent to algebraic computations when the residual loss is minimal, and effectively prevents oscillations in residuals caused by excessive fluctuations of turbulent viscosity during training. This method also allows us to directly initialize turbulent viscosity in the pre-training stage.

In this study, the effectiveness of the proposed framework is validated using a three-dimensional parameterized turbulent backward-facing step (BFS) flow problem.
The validation is performed using parameterized experiments conducted by Nadge et al. \cite{nadge2014high}, with Reynolds numbers $Re_h=6000\text{--}67000$, step expansion ratios $ER$=1.1, 1.3, 1.5, 1.84, and 2.50. 
As shown in Fig.~\ref{fig_1}a, The trained parameterized step flow prediction model includes two variable parameters: the geometric parameter step expansion ratio (ER) and the step Reynolds number ($Re_{h}$). 
The step expansion ratio $ER=(H+h)/H$ is controlled by varying the step height $h\in \left[0.4,2\right]\,\text{m}$ while keeping the inlet height $H=4\,\text{m}$ fixed. 
The Reynolds number $Re_{h}=Uh/\nu$ is controlled by varying the step height and fluid viscosity $\nu\in \left[0.00003,0.0004\right]\,\mathrm{m^2/s}$ while keeping the inlet velocity $U=(3,0,0)\,\text{m/s}$ fixed. 
The specific geometric proportions of the problem are consistent with the experiments, with the specific dimensions being $S=24\,\text{m}$, $L=52\,\text{m}$, and $L_2=40\,\text{m}$. 
The Reynolds number for the trained parameterized step flow prediction model ranges from $Re_{h} \in \left[3000,20,0000\right]$, and the step expansion ratio ranges from $ER\in \left[1.1,1.5\right]$. In this study, the symbol $Re$ represents $Re_h$.

The $k-\omega$ turbulence model \cite{wilcox1988reassessment} was employed to solve the parameterized turbulence problem. The Launder-Spalding wall function \cite{bredberg2000wall} was applied as wall boundary conditions and a velocity correction function was incorporated into the bottom wall boundary condition.
No additional data were introduced in the flow field to assist the training. In addition to experimental results, we also compared the model predictions with computational fluid dynamics (CFD) results. The CFD results were obtained by simulating the step flow field for each corresponding condition using the finite volume method in the Fluent software \cite{FLUENTtheory}. The detailed settings for PINNs and CFD simulations will be explained in the Methods section.

\begin{figure}
\refstepcounter{fig}
\includegraphics[width=0.95\linewidth]{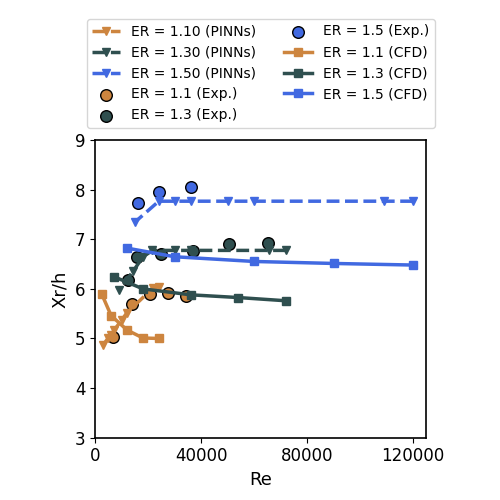}
\caption{\textbf{Comparison of PT-PINNs' reattachment length predictions with CFD and experimental results \cite{nadge2014high}.}  
}
\label{fig_3}
\end{figure}

\begin{figure*}[t]
\refstepcounter{fig}
\includegraphics[width=0.95\linewidth]{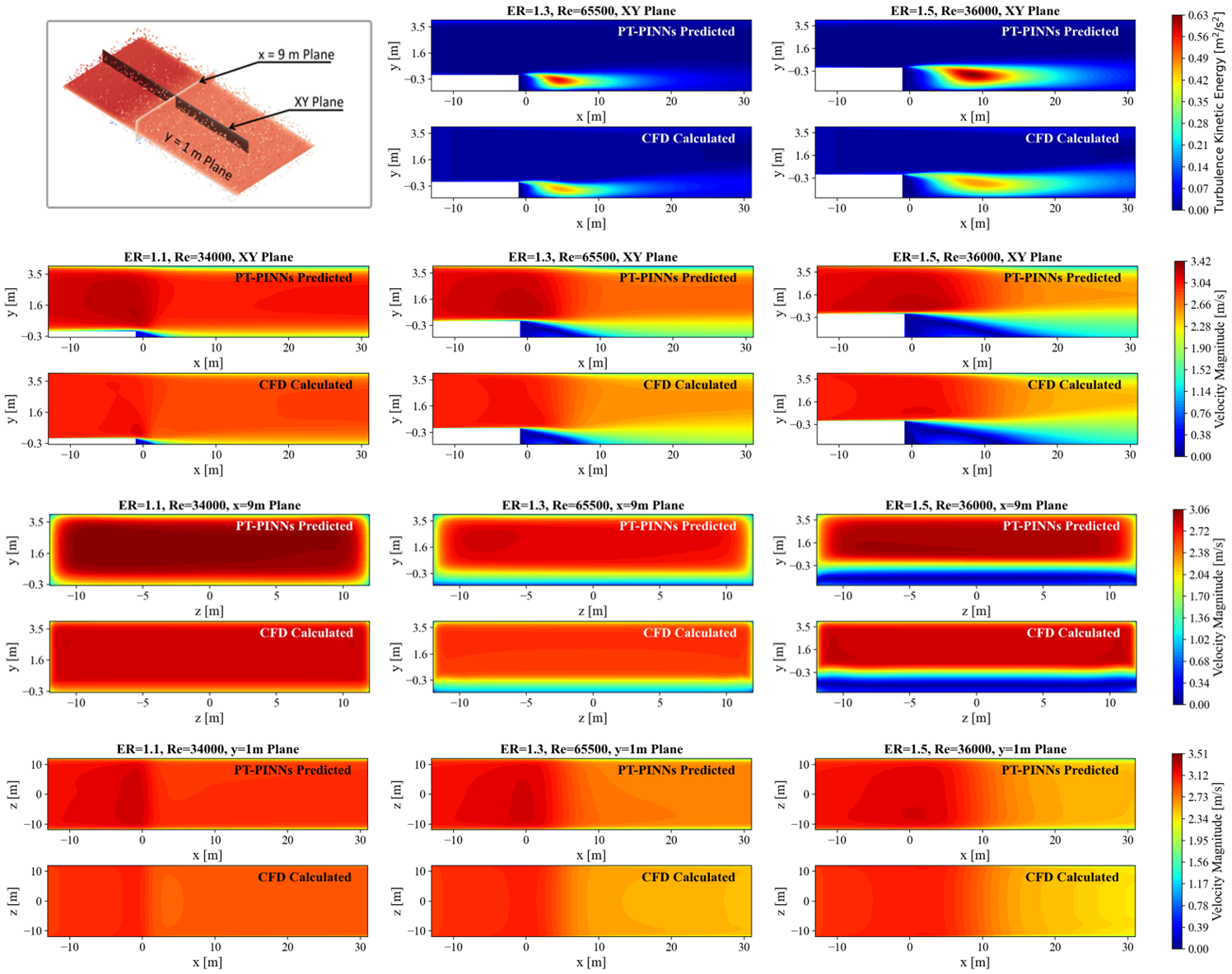}
\caption{\textbf{Comparisons of PT-PINNs predictions against CFD results of overall flow fields.}  
}
\label{fig_5}
\end{figure*}

\subsection*{Validation against Experiments and CFD Results}

Fig.~\ref{fig_4}a presents a comparison of the stream-wise velocity u along various sectional lines within the recirculation region, as predicted by PT-PINNs, with experimental \cite{nadge2014high} and CFD results under different operating conditions.
For the overall trend of velocity u variation along the sectional lines, PT-PINNs predictions achieve good agreements with both CFD and experimental results. 
Moreover, the predictions of PT-PINNs at the section lines of x/Xr = 0.15 and 0.45 exhibit even better alignment with experimental measurements than CFD results under various conditions. 
This demonstrates that PT-PINNs can provide predictions of recirculation vortex structures that align more closely with experimental results under varying operating conditions. 
Furthermore, it illustrates that PINNs are not inherently incapable of accurately predicting flow field details. 
By guiding PINNs to correctly learn the governing partial differential equations and boundary conditions through appropriate network architectures and training strategies, PINNs can achieve parameterized three-dimensional turbulent flow field predictions that meet industrial accuracy requirements.
Fig.~\ref{fig_4}b presents a comparison of the maximum turbulent stress $\mathrm{(u')^2/U_0^2}$ across vertical cross-sections at different $x$ position within the recirculation region, as predicted by PT-PINNs, with CFD and experimental results. 
Surprisingly, PT-PINNs achieve superior consistency with experimental data in terms of both the peak values and the overall distribution of maximum turbulent stress compared to CFD. 
This can be attributed to the utilization of automatic differentiation  \cite{baydin2018automatic} and the mesh-free calculation in PINNs, which enables a more detailed traversal of the flow field space through a random point sampling strategy in each iteration, thereby reducing the numerical errors when solving PDEs.

Fig.~\ref{fig_2} presents the comparisons of the flow field streamlines in the recirculation regions, as predicted by PT-PINNs, with experimental and CFD results across various step expansion ratios and Reynolds numbers.
Without the aid of internal field data for training, PT-PINNs accurately predicted the shapes of both the corner vortex and the recirculation vortex, demonstrating consistency across different expansion ratios and Reynolds numbers with experimental results and reflecting the self-similarity characteristics of turbulence.
Furthermore, as illustrated in Fig.~\ref{fig_3}, due to the application of refined boundary conditions based on experimental results, PT-PINNs can predict the reattachment length variation concerning the expansion ratio and Reynolds number more accurately, aligning more closely with experimental measurements than traditional CFD methods.
At high Reynolds numbers, the prediction of the reattachment length by CFD significantly deviates from the actual experimental measurements. 
This implies that while CFD can predict the overall shape of the recirculation vortex, the predicted vortex size differs considerably from experimental results. 
In contrast, the predictions of PT-PINNs show a high degree of agreement with the experimental results. 
This demonstrates the potential of PINNs as a parametric solver for achieving higher prediction accuracy in engineering turbulence problems compared to traditional CFD methods
Although we implemented a smooth transition in the loss function weights for the wall functions at the intersection of the horizontal and vertical walls of the step, minor discrepancies in PT-PINNs' flow field predictions persist near the upper horizontal wall of the step. These discrepancies can be attributed to conflicting velocity predictions from the wall functions of the two intersecting walls.

\begin{table*} 
\caption{\label{tab:table1} Comparisons of PT-PINNs and the CFD method.}
\centering
\scriptsize 
\begin{tabularx}{0.9\textwidth}{cccccccccc}
\toprule
\multicolumn{9}{c}{PT-PINNs training based on Nvidia Modulus platform \cite{MODULUSplatform}} \\ 
\midrule
Inputs & Outputs & \begin{tabular}[c]{@{}c@{}}Hidden layers \\$\times$ neurons\end{tabular} & \begin{tabular}[c]{@{}c@{}}Interior\\ batchsize\end{tabular} & Epoches & \begin{tabular}[c]{@{}c@{}}Time \\ costs (h)\end{tabular} & \begin{tabular}[c]{@{}c@{}}Inference speed \\ speed (s)\end{tabular} & GPU device & \begin{tabular}[c]{@{}c@{}} Model Size \\ (MB)\end{tabular}  \\ \midrule
x, y, z, $\nu$, ER & u, v, w, p, k, $\omega$ & $7\times512$  & 2048 & 400000 & 39 & 40 & 1$\times$RTX4090 GPU & 12.1 \\ 
\midrule
\multicolumn{9}{c}{CFD method based on ANSYS-Fluent software \cite{ANSYS}} \\ 
\midrule
Inputs & Outputs & \begin{tabular}[c]{@{}c@{}}Finite\\ cells\end{tabular}  & \begin{tabular}[c]{@{}c@{}}Simulation\\ iterations\end{tabular}  & \begin{tabular}[c]{@{}c@{}}Case\\ number\end{tabular} & \begin{tabular}[c]{@{}c@{}}Time \\ costs (h)\end{tabular} & \begin{tabular}[c]{@{}c@{}}Time costs \\ per case (h)\end{tabular} & CPU device &\begin{tabular}[c]{@{}c@{}}Data size \\ per case (GB)\end{tabular} \\ 
 \midrule
x, y, z & u, v, w, p, k, $\omega$ & 8000000 & 400 & $21^{2}$ & 632 & 1.5 & 1$\times$AMD EPYC 7b12 & 1.07  \\ 
 \midrule
\multicolumn{9}{l}{Note: The values of finite cells and Time costs per case are average values. The CPU and GPU devices are of equivalent procu-} \\
\multicolumn{9}{l}{-rement cost. The CPU device has 64 cores and is paired with 128 GB of memory.} \\ 
\bottomrule
\end{tabularx}
\end{table*}

Fig.~\ref{fig_5} presents the comparisons of the overall flow field predictions for the three-dimensional step flow by PT-PINNs and CFD methods under various conditions. 
On the $z=0\,\text{m}$, $y=1\,\text{m}$, and $x=9\,\text{m}$ plane the velocity predictions of PT-PINNs exhibit high consistency with the CFD results, with average relative errors in velocity magnitude of 5\%, 6\%, and 3\%, respectively. 
In terms of turbulence parameters, the turbulence kinetic energy predicted by PINNs, which are generated by the velocity gradient variations, also shows good agreement with the CFD results. 
Due to the application of the refined boundary condition based on experiment results \cite{nadge2014high, Jovi1995ReynoldsNE} at the bottom wall and the mesh-free nature of PINNs, the turbulence kinetic energy generated near the step is significantly higher than the CFD results but achieves better consistency with experimental results, as validated in Fig.~\ref{fig_4}b.

Table~\ref{tab:table1} presents the comparisons of the settings, solution time, and model size between PT-PINNs and the CFD methods for solving parameterized backward step flow.
Traditional CFD methods require multiple simulations to obtain the parameterized flow field distributions. If a lightweight parameterized prediction model is needed, training based on these simulation data is also required.
PT-PINNs, however, directly generate a lightweight neural network model capable of predicting parameterized turbulent proxy flow fields through a single training process, without the need for multiple simulations. Moreover, the model size is only 1\% of a single CFD case.
For the three-dimensional parameterized step flow case in this section, in terms of parameterized model construction time, the PT-PINNs method achieves an acceleration factor of 16 compared to CFD with multiple simulations (using a full-factorial parameter value design for 21×21 simulations).
Regarding the obtaining of flow fields with the same resolution for a single operating condition, the inference time of PT-PINNs is also only 1\% of that of a single CFD simulation.
This indicates that PT-PINNs hold promising prospects as a solver for parameterized engineering turbulent flow problems.

\subsection*{Turbulent Viscosity Constraint and Flow Fileds Initialization}

\begin{figure*}
\refstepcounter{fig}
\includegraphics[width=0.95\linewidth]{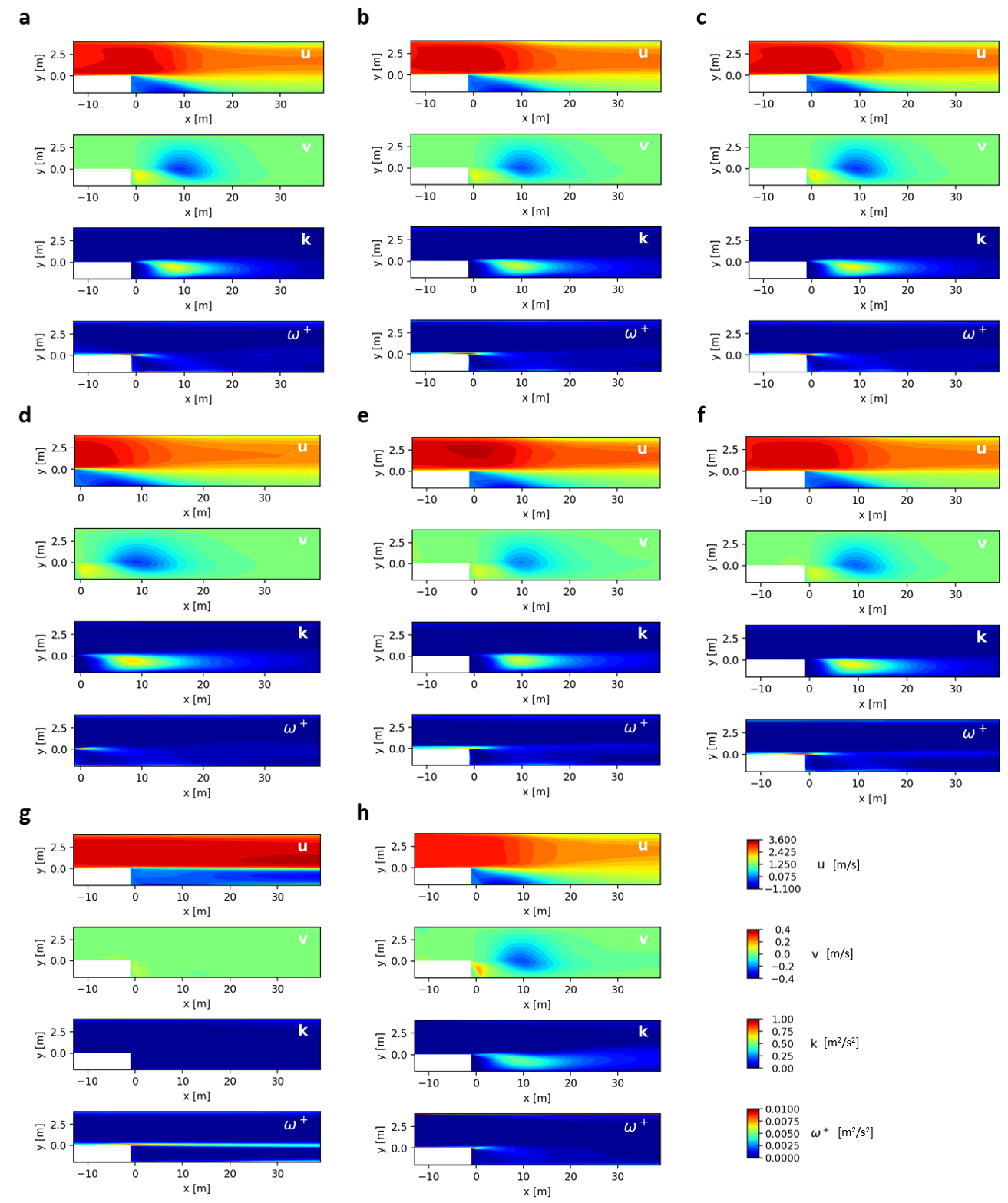}
\caption{\textbf{Comparisons of flow fields of $ER=1.5, Re=30000$ predicted by different networks under different weights balancing and pre-training methods.} \textbf{a.} The prediction of PT-PINNs trained with $\nu_t$ output network and pre-training under well-balanced weights. \textbf{b.} The prediction of PT-PINNs trained with $\nu_t$ output network and pre-training under poorly balanced weights. \textbf{c.} The prediction of PT-PINNs trained without using the pre-training method under well-balanced weights. \textbf{d.} The prediction of PT-PINNs trained without $\nu_t$ output network under well-balanced weights. \textbf{e.} The prediction of PT-PINNs trained without $\nu_t$ output network under poorly balanced weights. \textbf{f.} The prediction of PT-PINNs trained without $\nu_t$ output network or pre-training under well-balanced weights. \textbf{g.} The prediction of PT-PINNs trained without $\nu_t$ output network or pre-training under poorly balanced weights. \textbf{h.} CFD results.
}
\label{fig_6}
\end{figure*}

\begin{figure*}
\refstepcounter{fig}
\includegraphics[width=0.99\linewidth]{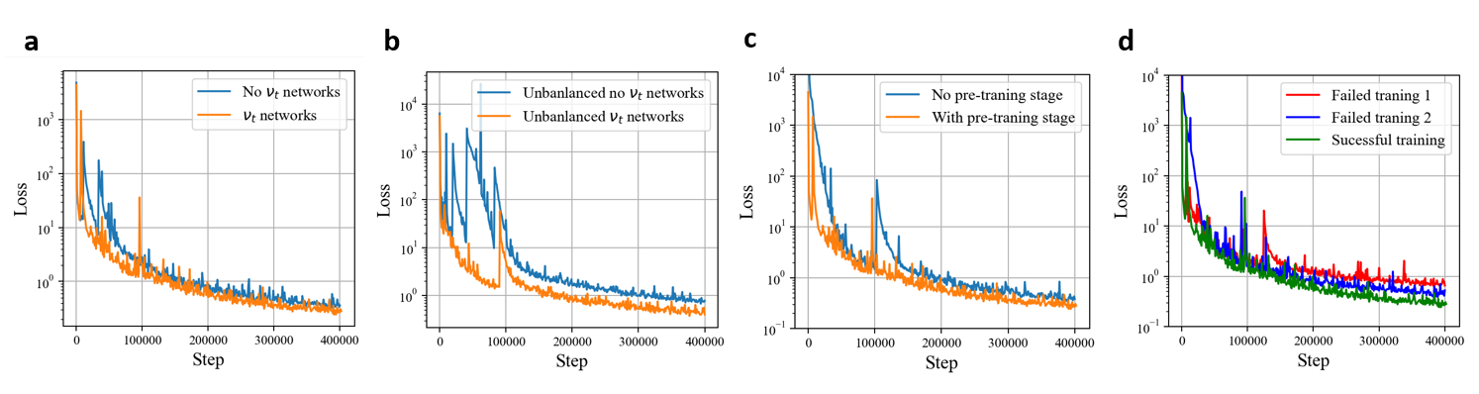}
\caption{\textbf{Comparisons of loss curves for different networks under different weights balancing and pre-training methods.} \textbf{a.} PT-PINNs with and without $\nu_t$ output network, trained with well-balanced weights. \textbf{b.} PT-PINNs with and without $\nu_t$ output network, trained with poorly balanced weights. \textbf{c.} PT-PINNs trained with and without the pre-training method under well-balanced weights. \textbf{d.} PT-PINNs with $\nu_t$ output network and pre-training method trained under well-balanced weights compared to PT-PINNs without the pre-training method, trained under poorly balanced weights (failed training 1), and PT-PINNs without $\nu_t$ output network, trained under poorly balanced weights(failed training 2).
}
\label{fig_7}
\end{figure*}

We developed two methods to enhance the PT-PINNs' ability to solve parameterized turbulent flow fields. 
The first method involves taking turbulent viscosity as a network output and transforming its algebraic expression into a soft constraint within the network architecture and training process. The second method is a pre-training method based on parameterized boundary conditions and flow field mass conservation. 
Fig.~\ref{fig_6} demonstrates the effects of these methods on PT-PINNs' parameterized turbulent flow field solving.
During the training process, we designed three distinct PDEs weight balancing schemes to further investigate the roles of these methods in avoiding local optima and reducing the sensitivity of model training to loss balancing. 
The three PDEs weight balancing schemes, corresponding to well-balanced weights, poorly balanced weights, and excessively imbalanced weights, are defined by the weight ratios of continuity-equation: momentum-equations: k-equation: $\omega^+$-equation as 1:10:0.1:1, 1:10:10:100, and 1:10:100:1000, respectively.
When the weights balance is well balanced and pre-training is applied, both neural networks with and without $\nu_t$ output achieve accurate overall flow field predictions (Fig.~\ref{fig_6}a and Fig.~\ref{fig_6}d). The relative errors of the velocity u compared to CFD results (Fig.~\ref{fig_6}h) were 7.5\% and 7.6\%, respectively.
However, under poorly balanced PDEs weights during training, the network without $\nu_t$ output produced significantly degraded predictions (Fig.~\ref{fig_6}e), with the u-velocity field error reaching 14.9\%, compared to 9.4\% for PT-PINNs with $\nu_t$ output trained under the same weights. Furthermore, the turbulent kinetic energy distribution near separation vortexes deviated significantly, resembling the trend observed in Fig.~\ref{fig_6}g for the network trapped in a local optimum.
Additionally, if the network with $\nu_t$ output is trained under a well-balanced weight but without pre-training applied, similar issues of the turbulent kinetic energy distribution arise. Its relative error of the velocity u increases to 10.5\%, whereas the error for the viscosity-output network without initialization remains low at 7.8\%. 
Fig.~\ref{fig_6}g illustrates the results of a network trapped in a local optimum due to imbalanced weight training. In this local optimum, the network sets turbulent kinetic energy to zero in regions where it should be high, achieving a physically unrealistic loss minimization.
When the weight balance is poorly balanced and pre-training is not applied or the weight balance is excessively imbalanced, the PT-PINNs' predictions exhibit the same issues.

Fig.~\ref{fig_7} illustrates the evolution of the total loss during training using different methods. PT-PINNs exhibit better convergence under both well-balanced and poorly balanced weights. 
However, PT-PINNs without the turbulent viscosity output method exhibit significantly greater oscillations during training with poorly balanced weights. 
Fig.~\ref{fig_7}c compares the loss curves of PT-PINNs trained with and without pre-training method. The former achieves faster and more robust convergence, requiring only 20,000 iterations without differential equation constraints (approximately 40 minutes), whereas the latter converges slowly.
Fig.~\ref{fig_7}d compares the iterative processes of networks trapped in local optima with those of a well-trained network. Networks trapped in local optima exhibit significantly higher converged losses compared to those avoiding local optima, though still within a relatively small range relative to the initial loss.

\section*{Discussion}
\label{sec3}

The parameterized prediction of turbulent flows is of significant importance in various engineering designs and digital twin scenarios. As a new partial differential equation solver leveraging neural networks, PINNs exhibit inherent advantages in addressing parameterized high-dimensional problems. 
In this study, we proposed a PINNs-based framework for solving parameterized steady-state turbulent problems, PT-PINNs. By incorporating turbulent viscosity as a network output and employing a flow-field-based pre-training method, we successfully enhance the prediction capability of PT-PINNs for engineering turbulence problems.

Using PT-PINNs, this study solved a parameterized three-dimensional backward-facing step flow problem based on the k-$\omega$ turbulence model without using any interior flow data for training.
We obtained a model that can quickly predict the parametric flow fields using only one training process by PT-PINNs.
The model includes two variable parameters, fluid viscosity, and step height, and can perform predictions for three-dimensional backward-facing step flows under arbitrary conditions with Re=3000-200,000 and ER=1.1-1.5. 
The model’s prediction capability was validated by comparing its results against Nadge's experimental data \cite{nadge2014high} and CFD results.
The model demonstrates good agreement with CFD simulation results in terms of the overall flow field, with relative velocity field errors kept below 6\%. 
For detailed comparisons in the recirculation regions, the shapes of the primary recirculation and corner vortexes closely match the experimental results. In some cases, the model's stream-wise velocity and turbulence stress predictions are even closer to experimental results than CFD simulations.

Regarding computational efficiency, training the model using the PT-PINNs framework took approximately 39 hours. While this duration is longer than a single CFD simulation, parameterized problems generally require hundreds or even thousands of CFD simulations. 
For instance, the PT-PINNs method achieves an acceleration factor of 16 compared to running multiple CFD simulations (using a full-factorial parameter design for 21×21 simulations). 
Furthermore, once trained, PT-PINNs can infer flow fields for any operating condition with an equivalent mesh resolution in just 40 seconds—only 1\% of the time required for a single CFD simulation.

Additionally, we validated the superiority of the two methods applied in the PT-PINNs framework for steady-state turbulent problems by solving the backward-facing step flow using the k-$\omega$ models with different methods.
Comparisons of different training methods show that the pre-training method and the method of treating turbulent viscosity as a network output help PT-PINNs maintain accurate predictions of turbulent flows, even under poorly balanced training weights.
In contrast, PT-PINNs without these two methods often produce poor predictions under various conditions and are more prone to converging to completely non-physical local optima.

\section*{Methods}
\addtocontents{toc}{\SkipTocEntry}
\subsection*{Turbulence Modeling}
\label{subsec2}

This work focuses on the steady-state incompressible Reynolds-averaged turbulence. The standard k-$\omega$ model \cite{wilcox1988reassessment} is employed to solve the parametrized three-dimensional turbulent backward-facing step flow problem in this study. The specific governing equations are as follows:
\begin{equation}
\frac{\partial u_i}{\partial x_i}=0 
\end{equation}
\begin{equation}
u_j \frac{\partial u_i}{\partial x_j}+\frac{1}{\rho} \frac{\partial p}{\partial x_i}-\frac{\partial}{\partial x_j}\left[\left(\nu+\nu_t\right)\frac{\partial u_i}{\partial x_j}\right]=0
\end{equation}
\begin{eqnarray} \label{equ-k}
u_j \frac{\partial k}{\partial x_j}=\frac{\partial}{\partial x_j}\left[\left(\nu+\frac{\nu_t}{\sigma_k}\right)\frac{\partial k}{\partial x_j}\right]+\frac{1}{\rho} (P_k - Y_k)
\end{eqnarray}
\begin{equation} \label{equ-om}
\begin{aligned}
\ u_i \cdot \frac{\partial \omega}{\partial x_i} = 
\frac{\partial}{\partial x_j}\left[\left(\nu+\frac{\nu_t}{\sigma_{\omega}}\right)\frac{\partial\omega}{\partial x_j}\right]-\frac{1}{\rho} (P_{\omega}-Y_ {\omega})
\end{aligned}
\end{equation}
where $u_i$ and $p$ are the Reynolds-averaged flow velocity and pressure, the density ${\rho}=1 \enspace \mathrm{kg/m^3}$ is constant. $\sigma_k=\sigma_{\omega}=2$ are model constants. $P_k=\rho v_t G$ represents the turbulent kinetic energy generation term.$\sqrt{G}$ represents the average strain rate tensor, and $G=2\left(u_x^2+v_y^2+w_z^2\right) +\left(u_y+v_x\right)^2+\left(u_z+w_x\right)^2+\left(v_z+w_y\right)^2 $. 
$Y_k=\rho \beta^* k \omega$ represents the turbulent kinetic energy dissipation term, where $\beta^* = 0.09$. $P_{\omega}=\rho \alpha G$ represents the turbulent specify dissipation rate generation term, where $\alpha = 5/9$. $Y_{\omega}=\rho \beta \omega^2$ represents the turbulent dissipation rate dissipation term, where $\beta = 0.075$.
The eddy viscosity ${\nu}_t$ is closed using the k-omega turbulence model, which is given by the following formula:
\begin{eqnarray}
\label{mut}
&\nu_t=\frac{k}{\omega}
\end{eqnarray}

Launder-Spalding wall functions \cite{bredberg2000wall} are utilized to solve the near wall flow.
The relations for Launder-Spalding wall functions formulation are similar to standard wall functions, except now the friction velocity can directly be computed from the turbulent kinetic energy as shown below:
\begin{equation}
u_\tau = C_\mu^{1 / 4} k^{1 / 2} 
\end{equation}
\begin{equation}
y^{+} = \frac{\rho u_\tau y_p}{\mu} 
\end{equation}
\begin{equation}
U = \frac{C_\mu^{1 / 4} k^{1 / 2}}{\kappa} \ln \left(E y^{+}\right) 
\end{equation}
\begin{equation}
\omega =(1-F)\frac{u_\tau}{0.3\kappa y} + F\frac{6\nu}{0.075y^2} 
\end{equation}
\begin{equation}
\left.\left.\tau_w \equiv \mu \frac{\partial U}{\partial y}\right|_w \approx\left(\mu+\mu_t\right) \frac{\partial U}{\partial y}\right|_P =\frac{\rho C_\mu^{1 / 4} k^{1 / 2} U \kappa}{\ln \left(E y^{+}\right)} 
\end{equation}
where $u_\tau$ is the friction velocity, $U$ is the mean velocity at the near wall region, $y^{+}$  is the dimensionless distance from the wall, $\tau_w$ is the wall shear stress. $y$ is the distance to the wall which is set as a constant in each wall constraint which is similar to \( y_p \) in CFD computations. In CFD computations, wall functions are applied to the first layer of grid cells near the wall. Therefore, the dimensionless parameter \( y^+ \) is calculated based on the height of the center of the first grid cell, \( y_p \). The height \( y_p \) needs to be set by estimating the turbulent boundary layer height according to the flow conditions. The function $F$ is introduced to smooth the $\omega$ calculations in different regions, which is defined as: $F=(tanh(10 - y^+) + 1.0) ^{0.5}$.

When using wall functions for turbulent flow simulations, the turbulent flow in the separation and recirculation regions often can not be accurately predicted. 
As shown in Fig.~\ref{fig_3}, the predicted reattachment length using the wall-function-based CFD method demonstrated poor agreement with experimental results across the parameter space.
Based on the summaries of the wall shear stress and reattachment point distribution from Nadge et al. \cite{nadge2014high} and Jovic and Driver \cite{Jovi1995ReynoldsNE} for backward-facing step flows, a set of empirical formulas was employed to modify the wall functions velocity for the boundary conditions at the bottom and vertical walls:
\begin{equation}
v_{ve}= 0.07sin(2\pi y^2/h^2-\pi) 
\end{equation}
\begin{equation}
    u_{bo} =
\begin{cases}
0.025\cdot(erf(16(0.5 + (x - 0.6h) / h)) \\
+ erf(16 (0.5 - (x - 0.6h) / h))), \\
\quad \quad \quad \quad  \ \text{if } 0 \leq x < 1.2h, \\
0.65 \cdot sin(\pi(x-Xr)/(Xr-h)), \\
\quad \quad \quad \quad  \ \text{if } 1.5h \leq x < Xr + 1.2h,\\
1.76ER \quad  \text{if } Xr + 8h \leq x ,
\end{cases}
\end{equation}
\begin{equation}
Xr=c_{re}\cdot(4.95 * ER  + 0.5)\cdot h \quad \quad
\end{equation}
\begin{equation}
c_{re} = Min(1, 1 - 0.000011*(20000-Re_h))
\end{equation}
where $x$ and $y$ are the horizontal and vertical distance to the downstream step corner, and $X_r$ is the reattachment length. $v_{ve}$ is the wall boundary velocity of the vertical wall, $u_{bo}$ is the wall boundary velocity of the bottom wall.

\subsection*{Physics Informed Neural Networks}
\label{subsec3}
This work employs the Physics-Informed Neural Networks (PINNs) method as a parameterized solver to address the parameterized flow problem. Its input $X$ includes spatial coordinates as well as arbitrary geometric, operating conditions, and physical property parameters relevant to the problem. 
The output $Y$ represents the flow field variables, and the relationship between them is established through a fully connected neural network, which can be mathematically expressed as follows:
\begin{equation}
H_1 = \sigma(\bm{W}_1 \bm{X} + \bm{b}_1) 
\end{equation}
\begin{eqnarray}  
\bm{H}_l = \sigma(\bm{W}_{l-1} \bm{H}_{l-1} + \bm{b}_l) \quad \quad \quad \nonumber \\ 
\quad \quad \quad \quad \quad \quad \quad \quad for \ l=2,3,...,n
\end{eqnarray}
\begin{equation}
\bm{Y}(\bm{X}) = \bm{W}_{n+1} \bm{H}_n + \bm{b}_{n+1}
\end{equation}
where, $\bm{W}$ and $\bm{b}$ are the network weights and biases, $n$ is the number of hidden layers, and $\sigma$ is the activation function. 
The input $\bm{X}$ is mapped through the network with multiple hidden layers $\bm{H}_l$, deriving the relationship between the variables in $\bm{Y}$ and the parameters in $\bm{X}$. 
During the iterative process, automatic differentiation is used \cite{baydin2018automatic} to calculate the derivatives of the output $\bm{Y}$ with respect to $\bm{X}$, enabling the solution of differential equations in the physics loss function. 
The iterative correction of weights and biases can be viewed as fitting a specific function that complies with physical constraints in the parameterized space.
In this work, the loss functions are expressed as:
\begin{equation}
    L_{\text{total}} = L_{\text{PDE}} + L_{\text{BC}} 
\end{equation}
\begin{eqnarray}  
    L_{\text{PDE}} = w_1 L_{continuity} + w_2 L_{moumentum}  \nonumber \\
    + w_3 L_{k-transpot} + w_4 L_{\omega-transpot} 
\end{eqnarray}
\begin{equation}
    L_{\text{BC}} = w_5 L_{inlet} + w_6 L_{outlet} + w_7 L_{wall-function} 
\end{equation}
\begin{equation}
    L_{\nu_t} = w_8 ||(\nu_t \omega -k)||^2 + w_9|| (\nu_t-\frac{k}{\omega})||^2
\end{equation}
where $w_i$ represents the weights of each residual in the loss functions.
The turbulence viscosity constraint starts with a weaker constraint in the initial steps and transitions to a stronger one later. The transfer is defined as follow:
\begin{equation}
    w_{8}=(tanh((120000-n_{step})\times5\times10^{-5})+1)\times0.5 
\end{equation}
\begin{equation}
    w_{9}=1-w_{8}
\end{equation}
where $n_{step}$ is the iteration step number during training.
Since the presence of $\omega$ in the denominator in the constraint of $\nu_t=\frac{k}{\omega}$ often causes significant oscillations in the loss function during training, this strategy enables the neural network to first learn the overall flow field by initially incorporating a weaker form of the constraint, then progressively refining the detailed flow field through a gradual transition to the stronger form.

In this study, most other weights of loss functions are specified constant values. At the step's vertical wall, we applied a smooth weight transition to minimize the influence of corner points on training:
\begin{eqnarray}  
    w_{\text{normal-vel}} = (\text{erf}(64 \cdot (0.5 + (y + h/2) / 0.9h ))) & \nonumber \\
     + \text{erf}(64 \cdot (0.5 - (y + h/2) / 0.9h))) & 
\end{eqnarray}

A pre-training step is introduced as flow field initialization before formal training. In this step, the governing partial differential equation constraints are not applied to the interior field. 
For the velocity field, specific regions are directly constrained using constant values based on mass conservation, while other regions used integrated interfaces to maintain mass conservation. 
Boundary conditions are applied normally in the pre-training stage. For other physical quantities, constant values are directly assigned. For the BFS flow case in this study, the loss functions of pre-training stage can be expressed as:
\begin{eqnarray}  
    L_{init}=w_{10}L_{\text{interior-1, init-}u} + w_{11}L_{\text{interior-2, init-}u}  \nonumber \\ 
    + w_{12}L_{\text{interior, init-}v}+ w_{13}L_{\text{interior, init-}w}  \nonumber \\ 
     + w_{14}L_{wall-function} +  w_{15}L_{\text{interior, init-}\nu_t}  
\end{eqnarray}
\begin{equation}
    L_{\text{interior-1, init-}u} = u - u_{\text{inlet}}
\end{equation}
\begin{equation}
    L_{\text{interior-2, init-}u} = u - u_{\text{inlet}} / ER 
\end{equation}
\begin{equation}
    L_{\text{interior, init-}v} = v 
\end{equation}
\begin{equation}
    L_{\text{interior, init-}w} = w 
\end{equation}
\begin{equation}
    L_{\text{interior, init-}\nu_t} = \nu_t - 0.003 
\end{equation}
where interior-1 represents the fluid region upstream of the step, while interior-2 denotes the fluid region located $x>10h$ downstream of the step corner. To ensure flow rate conservation within the fluid region between these two areas, we apply several integral sections.

The Adaptive Moment Estimation (Adam) optimizer \cite{kingma2014adam} was used to adjust the parameters of the neural network model. The initial learning rate is set to 0.0001, and it decays to 95\% of its original value every 8,000 steps. The training is repeated iteratively until the loss function reaches the user-defined threshold.

\bibliography{name}

\section*{Acknowledgment} 
This work was supported in part by the National Natural Science Foundation of China under Grand 52236002 (K.L.(Kun Luo)).

\section*{Author Contributions}
L.J.(Liang Jiang) was responsible for conceptualization, methodology design, conducting numerical experiments, and drafting the original manuscript. Y.C.(Yuzhou Cheng) and K.L.(Kun Luo) extensively reviewed and edited the manuscript, providing valuable suggestions and revisions. Y.C. also contributed to methodology discussions. J.F.(Jianren Fan) and K.L. supervised the work

\section*{Competing Interests}
The authors declare no competing interests.

\end{document}